\documentclass[aps,english,prc,preprintnumbers,fleqn,showkeys,
superscriptaddress,floatfix]{revtex4}

\usepackage[latin9]{inputenc}
\usepackage{wrapfig}
\usepackage{calc}
\usepackage{amsmath}
\usepackage{graphicx}
\usepackage{float}
\usepackage{calc}
\usepackage{amstext}
\usepackage{natbib}
\usepackage{epsf}
\def\beq{\begin{equation}}
\def\eeq{\end{equation}}
\def\bea{\begin{eqnarray}}
\def\eea{\end{eqnarray}}
\def\beqa{\begin{equation}\begin{array}{l}}
\def\eeqa{\end{array}\end{equation}}
\def\figlab#1{\label{fig:#1}}

\def\figref#1{fig.~\ref{fig:#1}}

 \def\De{\Delta}

\def\lag{{\mathcal L}}

\def\mathscr{\mathcal}

\def\3d{3-D}

\def\ol#1{\overline{#1}}
\def\amm{a.m.m.}

\def\s#1{\setbox0=\hbox{$#1$}  % set a box for #1
   \dimen0=\wd0     % and get its size
   \setbox1=\hbox{/} \dimen1=\wd1  % get size of /
   \ifdim\dimen0>\dimen1   % #1 is bigger
      \rlap{\hbox to \dimen0{\hfil/\hfil}} % so center / in box
      #1     % and print #1
   \else     % / is bigger
      \rlap{\hbox to \dimen1{\hfil$#1$\hfil}} % so center #1
      /      % and print /
   \fi}      %

\begin{document}

\title{Electromagnetic properties of baryons}

%\classification{14.20.-c,13.40.Em  }


\author{T. Ledwig}
\affiliation{Institut f\"ur Kernphysik, Universit\"at Mainz, D-55099 Mainz, Germany}

\author{J. Martin-Camalich}
\affiliation{Department of Physics and Astronomy, University of Sussex, BN1 9Qh,
    Brighton, UK.\\
Departamento de Fisica Teorica and IFIC, Universidad de Valencia-CSIC, Spain}

\author{V. Pascalutsa}
\affiliation{Institut f\"ur Kernphysik, Universit\"at Mainz, D-55099 Mainz, Germany}

\author{M. Vanderhaeghen}
\affiliation{Institut f\"ur Kernphysik, Universit\"at Mainz, D-55099 Mainz, Germany}

%\author{T. Ledwig}{ address={Institut f\"ur Kernphysik, Universit\"at Mainz, D-55099 Mainz, Germany}}
%\author{J. Martin-Camalich}{  address={Departamento de Fisica Teorica and IFIC, Universidad de
%    Valencia-CSIC, Spain\\
%    Department of Physics and Astronomy, University of Sussex, BN1 9Qh,
%    Brighton, UK.}}
%\author{V. Pascalutsa}{ address={Institut f\"ur Kernphysik, Universit\"at Mainz, D-55099 Mainz, Germany}}
%\author{M. Vanderhaeghen}{address={Institut f\"ur Kernphysik, Universit\"at Mainz, D-55099 Mainz, Germany}}

\begin{abstract}
We discuss the chiral behavior of nucleon and $\Delta(1232)$
electromagnetic properties within the framework of a $SU(2)$ covariant baryon chiral
perturbation theory. Our one-loop calculation is complete to the
order $p^3$ and $p^4/\Delta$ with $\Delta$ as the $\Delta(1232)$-nucleon
energy gap. We show that the magnetic moment of a resonance can be defined through
the linear energy shift only when an additional relation between the involved
masses and the applied magnetic field strength is fulfilled. Singularities and
cusps in the pion mass dependence of the
$\Delta(1232)$ electromagnetic moments reflect a non-fulfillment.
We show results for the pion mass dependence of the nucleon iso-vector electromagnetic
quantities and present results for finite volume
effects on the iso-vector anomalous magnetic moment.
\end{abstract}

\maketitle

\section{Anomalous magnetic moment of a resonance}

We consider the energy shift $\Delta E(\vec{B})$ of an unstable spin-1/2 particle $\Psi$ of mass
$M_*$ in an external magnetic field $\vec{B}$. The decay products are a spin-1/2 particle $\psi$ and a scalar particle
$\phi$ of masses $M$ and $m$, respectively. We take a Yukawa interaction of
strength $g$ given by: $\lag_{\mathrm{int}} = g\, \Big(\, \ol{\Psi}
\,\psi\,\phi  + \ol \psi \,{ \Psi}\, \phi^\ast \Big) $. For stable particles
with spin $s_z=+1/2$ and $\vec{B}=Be_z$, we can normally write the energy shift induced
by the anomalous magnetic moment (\amm) $\kappa_*$ as $\De \tilde{E}=\frac{\De
  E}{M_\ast} + \frac{1}{2} \tilde B= -\frac{\kappa_*}{2} \tilde B  $ with
$\tilde{B}=e B_z/M_\ast^2$. This is not generally possible
for unstable particles \cite{QSparticles}. To see this, we calculate $\De \tilde{E}$ by the
background field method of \cite{Sommerfield:electronAMM}:
\beq
\De \tilde{E} = \frac{ g^2}{(4\pi)^2}\int_0^1 \! dx \,(r+x) 
\ln\Big[1+\frac{ x (1-x)\, \tilde{B}}{x \mu^2 - x (1-x) + (1-x) r^2}\Big]\,\,\,,
\eeq
with $r=M/M_\ast$, $ \mu=m/M_\ast$. After integrating the Feynman parameter,
we obtain contributions which are non-analytic in $\tilde{B}$. These terms can only be
expanded, i.e. a magnetic moment defined by $\Delta E= -\vec{\mu}\cdot\vec{B}$, when the following condition is met:
\beq
\label{cond1}
|eB|/2M_*\ll|M_*-M-m|\,\,\,.
\eeq
A mass of $m=M_*-M$ violates the above condition which is reflected by
singularities and cusps in electromagnetic (em) moments. Examples are the
em moments of the $W$-boson \cite{WBoson}. The loop contributions with
top and bottom quarks are singular when the masses are tuned to $m_t = m_W -  m_b$. Another example are the
em moments of the $\Delta(1232)$. We obtain singularities and cusps in the
pion mass dependence of the $\Delta(1232)$ magnetic dipole, electric quadrupole and magnetic
octupole moments at $m_\pi = M_\Delta - M_N$ \cite{DeltaCHIRALbehavior}. 

\section{Nucleon electromagnetic properties}

The nucleon Dirac $F_1(Q^2)$ and Pauli $F_2(Q^2)$ form factors are defined by
the following matrix element:
\begin{equation}
\langle
N(p^{\prime})|\overline{\Psi}(0)\gamma^{\mu}\Psi(0)|N(p)\rangle=\overline{u}(p^{\prime})\left[\gamma^{\mu}F_{1}(Q^{2})+\frac{i\sigma^{\mu\nu}q_{\nu}}{2m_{N}}F_{2}(Q^{2})\right]u(p)\,\,\,,\label{Eq:NucleonFF0}
\end{equation}
with $q=p^{\prime}-p$, $Q^{2}=-q^{2}$ as the momentum transfer and
$u(p)$ as the nucleon spinor with mass $m_{N}$. At $Q^2=0$ these form factors
are the nucleon charge $e_{N}$, anomalous magnetic 
moment $\kappa_{N}$ and magnetic moment $\mu_N$: $F_{1}(0)  =  e_{N}$, $F_{2}(0)  =
\kappa_{N}$ and $\mu_N = (e_N + \kappa_N) e/2M_N$.

We concentrate on the chiral behavior of the nucleon iso-vector anomalous magnetic
moment $\kappa_V$, Dirac $\langle r^{2}_1\rangle_V$ and Pauli $\langle
r^{2}_2\rangle_V$ radii that are defined by:
\begin{equation}
\kappa_V = F^p_2(0)-F^n_2(0)\;\;\;\;,\;\;\;\;\langle r^{2}_i\rangle_V=-\frac{6}{F_i^V(0)}\frac{dF_i^V(Q^{2})}{dQ^{2}} \,\,\, .
\end{equation}
We use the chiral perturbation theory given by the B$\chi$PT Lagrangian of \cite{Gasser(1988):ChPT}
with the $\Delta(1232)$-isobar included by the $\delta$-power
counting scheme of \cite{Pascalutsa(2003):deltaPOWERCOUNTING}. The explicit
Lagrangian consisting of pion, nucleon, $\Delta(1232)$-isobar and photon
fields can be found in \cite{Pascalutsa(2005):DeltaMDM,DeltaLagrangian}. The
power counting breaking terms as found in \cite{Gasser(1988):ChPT} are
treated by the renormalization prescription of \cite{Gegelia(1999):EOMS}. Our B$\chi$PT investigation is
complete to the order $p^3$ with inclusion of $\Delta(1232)$ effects up to
$p^4/\Delta$ where $\Delta$ is the $\Delta(1232)$-nucleon energy gap. 

In \figref{fig:DiagramsNucleon} we depict all one-loop contributions
originating at the considered order. For the diagram (N4) we
also investigate contributions from a non-minimally coupled photon to the
$\Delta \Delta \gamma$ vertex as discussed in \cite{Alexandrou(2009):LatticeDelta}. 
\begin{figure}[h]
\includegraphics[scale=0.25]{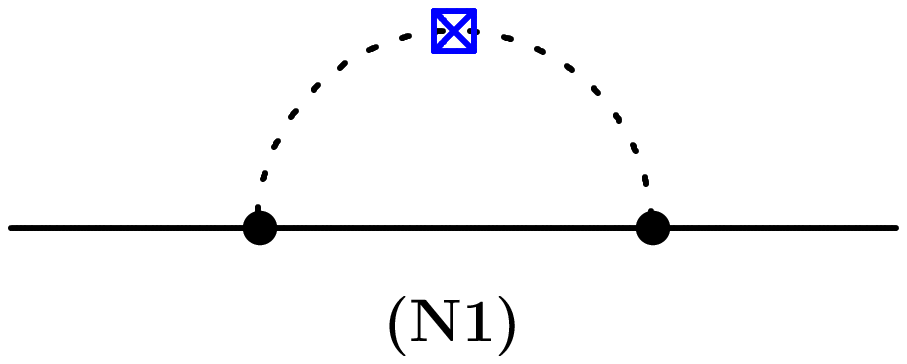}\includegraphics[scale=0.25]{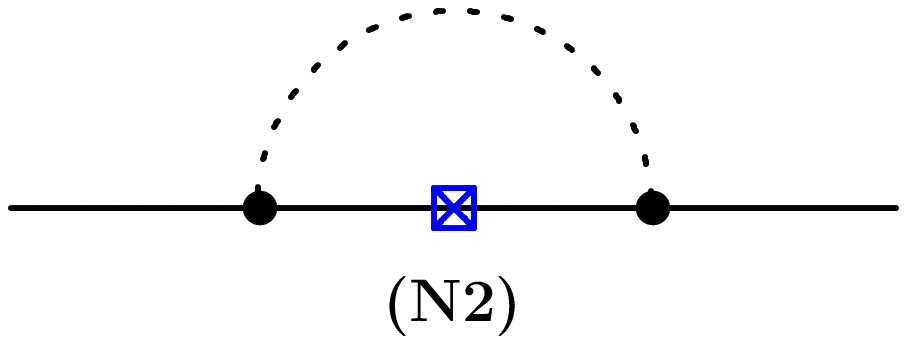}\includegraphics[scale=0.25]{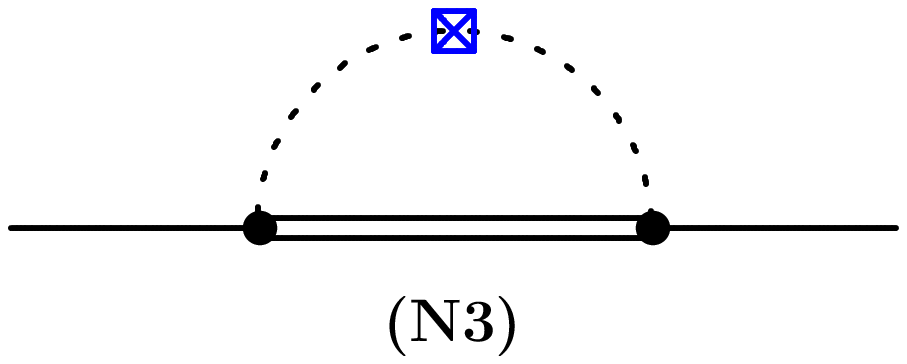}\includegraphics[scale=0.25]{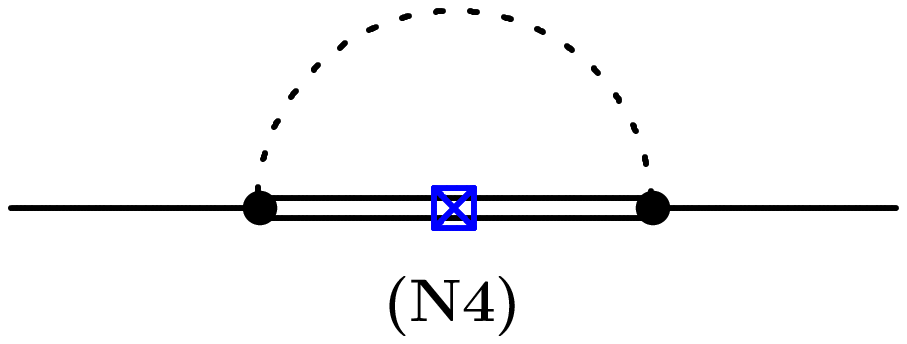} 
\end{figure}
\begin{figure}[h]
\includegraphics[scale=0.25]{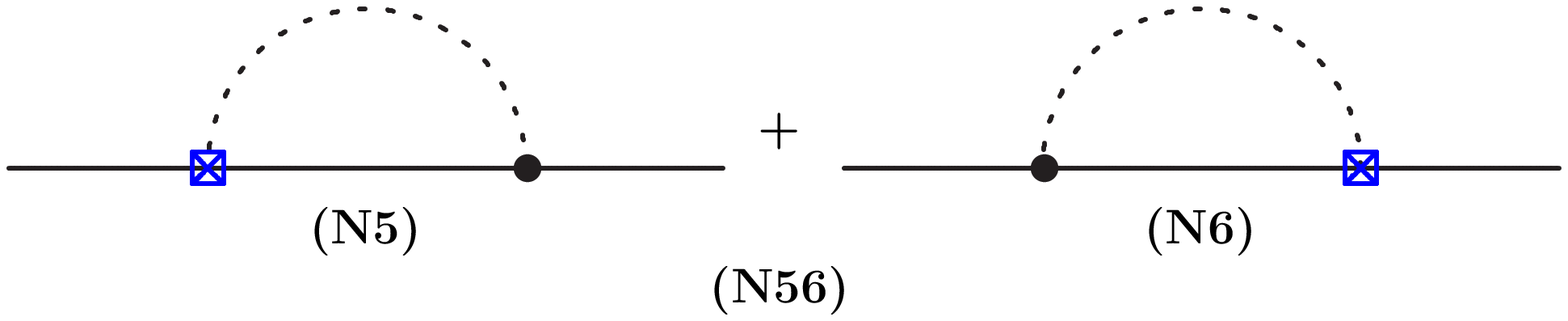} \includegraphics[scale=0.25]{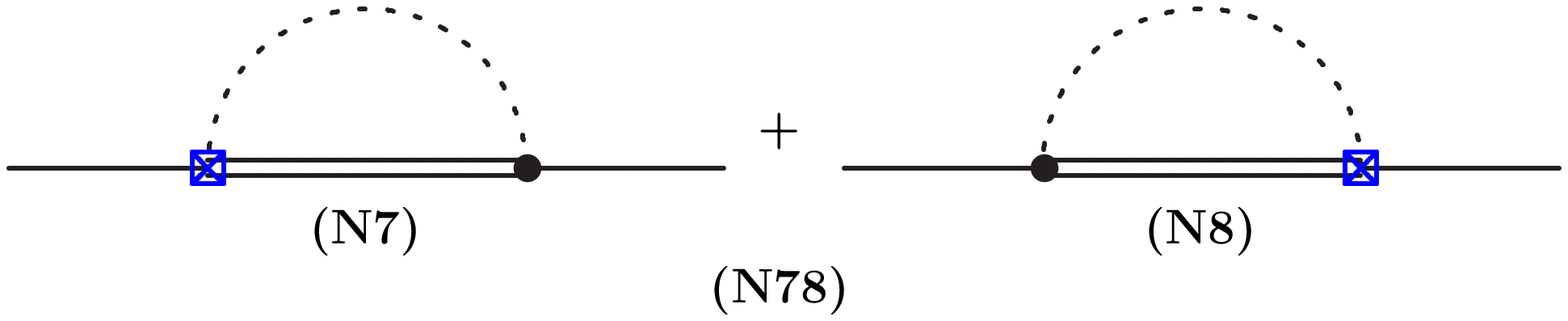}
\caption{\figlab{fig:DiagramsNucleon} Diagrams contributing to the nucleon electromagnetic properties.
Solid-single lines represent nucleons, solid-double lines the $\Delta(1232)$-isobar
and dashed lines the pions. The photon coupling is denoted by the
blue squares while the $N N \pi$ and $\Delta N \pi$ vertices by black dots. For
the Dirac radius we also include the tadpole graphs that are not depicted here.}
\end{figure}
\begin{figure}[h]
\includegraphics[scale=0.175]{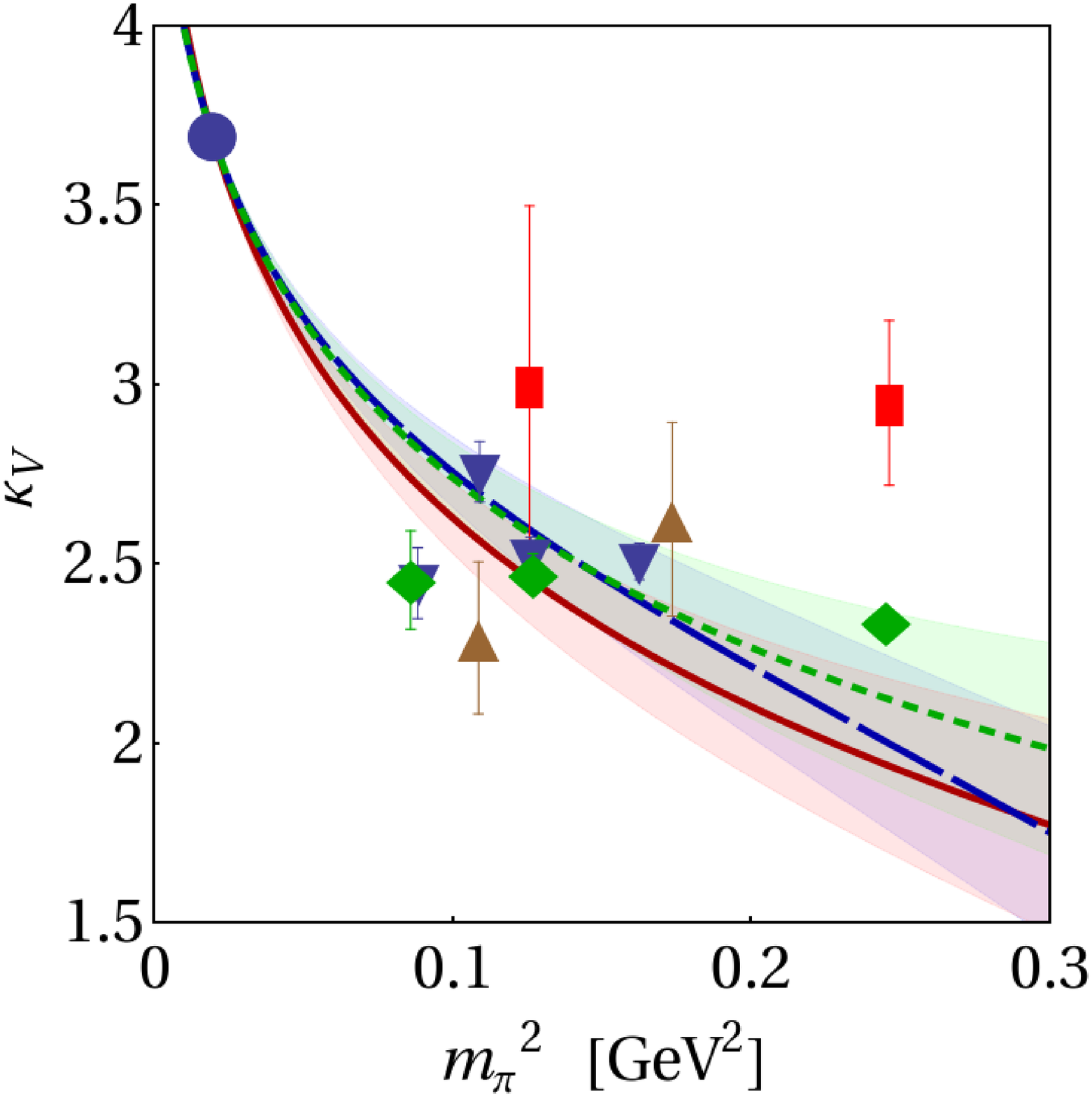}\;\;\hspace{0.75cm}\includegraphics[scale=0.175]{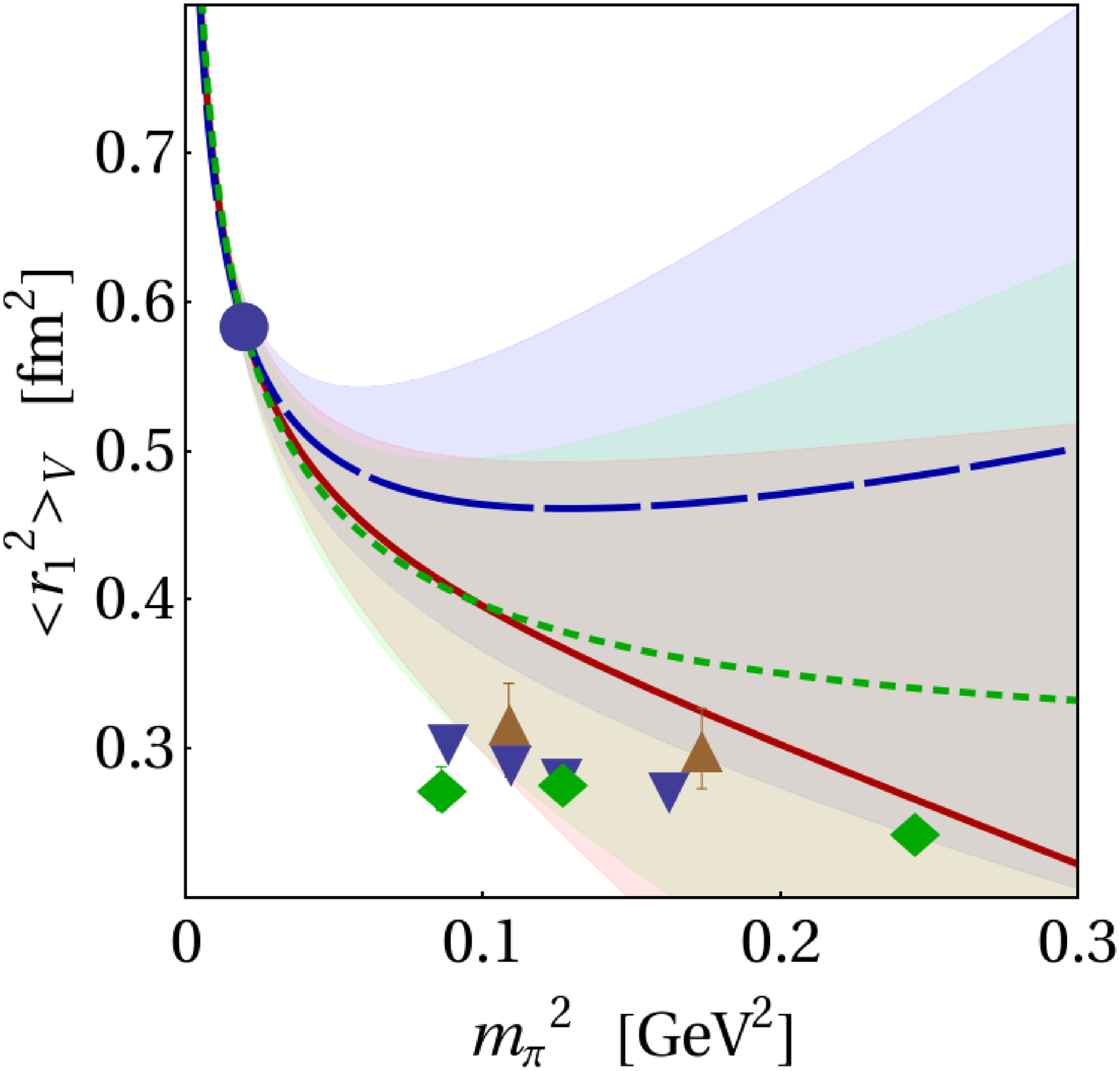}\;\;\hspace{0.75cm}\includegraphics[scale=0.175]{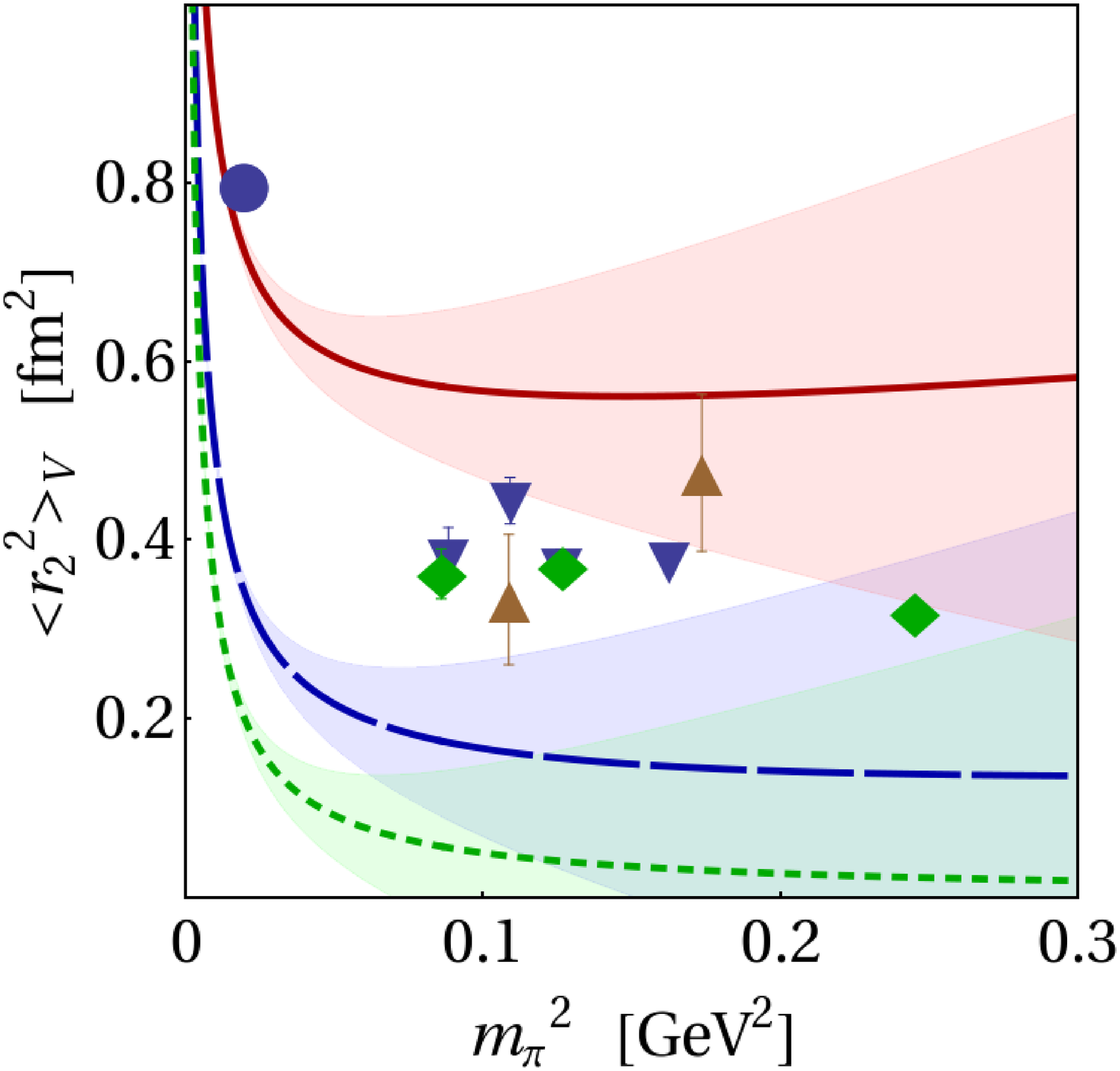}
\caption{\figlab{fig:NucleonPlots} Nucleon iso-vector anomalous magnetic
  moment $\kappa_V$, Dirac $\langle r_1^2 \rangle_V$ and Pauli $\langle r_2^2
  \rangle_V$ radii as function of the pion mass squared. The lattice data are
  taken from: brown up-triangles \cite{Yamazaki}, red rectangles \cite{Lin}, blue down-triangles
  \cite{Syritsyn}, green squares \cite{Bratt}. The green dotted line corresponds to
  contributions coming from the graphs (N1) and (N2). The red solid line corresponds to the contributions of
  all graphs with a non-minimal $\Delta$-photon coupling in (N4)
  while the blue dotted line is with a minimal $\Delta$-photon coupling. We
  constrain our results for $\kappa_V$ and $\langle r_1^2 \rangle_V$ to the phenomenological
  values $\kappa_V=3.7$ and $\langle r_1^2 \rangle_V= (0.765)^2\,\textrm{fm}^2$
  at $m_\pi=139$ MeV which are denoted by blue circles.}
\end{figure}

In \figref{fig:NucleonPlots} we show our results \cite{DeltaCHIRALbehavior} for the pion mass
dependence of the nucleon iso-vector \amm, Dirac $\langle r_1^2 \rangle_V$
and Pauli $\langle r_2^2 \rangle_V$ radii. To the order $p^3$ there appear
counter terms for the \amm~and $\langle r_1^2 \rangle_V$ while one for the
$\langle r_2^2 \rangle_V$ appears at higher orders. 

\section{Finite volume effects for $\kappa_V$}

We now discuss preliminary results for finite volume contributions
to the nucleon iso-vector \amm~coming from the graphs (N1) and (N2) of
\figref{fig:DiagramsNucleon}. For this, we calculate the self-energy of
the nucleon in a constant magnetic field $\vec{B}=Be_z$. We take here the
pseudo-scalar $NN\pi$ vertex. We obtain for the self-energies of the proton and neutron the following results:
\beq
\label{eq:sigma}
\left.\begin{array}{c}
\begin{array}{ccc}
\Sigma_p & = & 2\Sigma_1+\Sigma_2\\
\Sigma_n & = & 2\Sigma_3\end{array}\end{array}\right\}
\;\;\textrm{with}\;\;\;\Sigma_i = \frac{g^2m_N}{i(4\pi)^4} \int_0^1dz \int
d^4l \frac{1-z}{[l^2+m_\pi^2z-(1-z)^2m_N^2-\beta_i)]^2}\,\,\,
\eeq
and $\beta_1=z(1-z)B$, $\beta_2=-(1-z)^2B$, $\beta_3=-(1-z)B$. In a finite
volume the loop momentum $\vec{l}$ is discretized by
$\int\frac{d^{3}\vec{l}}{(2\pi)^{3}}\to\frac{1}{L^{3}}\sum_{\vec{n}}$ with
$\vec{l}=\frac{2\pi}{L}\vec{n}$ and $\vec{n}\in Z^3$. We extract the
\amm~through the linear coefficient in an expansion in $B$. The \amm~infinite
 volume contribution obtained from \eqref{eq:sigma} 
coincides with the one discussed in the previous section.

In \figref{fig:FV} we depict our results for the nucleon iso-vector \amm~for a
volume of size $L=3$ fm with $n_i\leq1$ and $n_i\leq 2$. The main
effect comes from the $n_i\leq1$ contribution which adds negatively to the infinite
volume result.

\begin{figure}[h]
\includegraphics[scale=0.48]{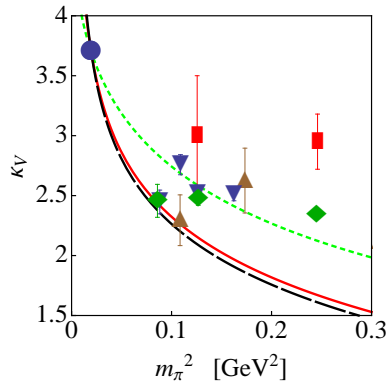}
\caption{\figlab{fig:FV} Nucleon iso-vector anomalous magnetic moment. The
  lattice data is the same as in \figref{fig:NucleonPlots}. The dotted green line
  is our infinite volume result obtained from \eqref{eq:sigma}
  which coincide with the dotted green line of
  \figref{fig:DiagramsNucleon}. The other two lines correspond to the finite volume
  contributions for a box size of $L=3$ fm and $n_i\leq1$ for the red line and
  $n_i\leq 2$ for the dashed black one.}
\end{figure}

\section{Summary}
We first discussed the anomalous magnetic moment of a resonance. In contrast
to stable particles, the self-energy of unstable particles depends
non-analytically on the applied external magnetic field. A resonance magnetic moment is
defined by the linear energy shift $\Delta E = -\vec{\mu}\cdot\vec{B}$ only when condition \eqref{cond1} is
met. We calculated the pion mass dependence of the $\Delta(1232)$ electromagnetic
moments which reveal singularities and cusps for the pion mass at which
\eqref{cond1} is not fulfilled.
We studied the chiral behavior of the nucleon iso-vector \amm, Dirac-
and Pauli-radii in a manifestly covariant $SU(2)$ baryon chiral perturbation
theory with inclusion of $\Delta(1232)$-isobar degrees of freedom. Our results
are complete to the order $p^3$ and $p^4/\Delta$. We also showed preliminary
results for finite volume effects on the iso-vector \amm~.

\section*{Acknowledgments}
%\begin{theacknowledgments}
The work of T.L. was partially supported
by the Research Centre Elementarkraefte und Mathematische
Grundlagen at the Johannes Gutenberg University Mainz. JMC acknowledges the MEC contract FIS2006-03438, the EU Integrated Infrastructure Initiative Hadron Physics Project contract RII3-CT-2004-506078 and the Science and Technology Facilities Council [grant number ST/H004661/1] for support.
%\end{theacknowledgments}


\begin{thebibliography}{}

\bibitem{QSparticles} T. Ledwig, V. Pascalutsa, M. Vanderhaeghen, Phys. Rev. D {\bf82}, 091301 (2010).

\bibitem{Sommerfield:electronAMM} C. M. Sommerfield, Annals of Physics {\bf5}, 26 (1958).

\bibitem{WBoson} G. Couture, J. N. Ng, Z. Phys. C {\bf35}, 65 (1987).

\bibitem{DeltaCHIRALbehavior} T. Ledwig, J. Martin-Camalich, V. Pascalutsa,
  M. Vanderhaeghen, in preparation.  

\bibitem{Pascalutsa(2005):DeltaMDM} V. Pascalutsa and M. Vanderhaeghen,
Phys. Rev. Lett. \textbf{94}, 102003 (2005).

\bibitem{Gasser(1988):ChPT}J. Gasser, M. E. Sainio, A. Svarc, Nucl.
Phys. B \textbf{307}, 779 (1988).

\bibitem{Pascalutsa(2003):deltaPOWERCOUNTING} V. Pascalutsa and D.
R. Phillips, Phys. Rev. C \textbf{67}, 055202 (2003). 

\bibitem{DeltaLagrangian} V. Pascalutsa, M. Vanderhaeghen,
Phys. Rev. D \textbf{73}, 034003 (2006); Phys. Rev. D \textbf{77}, 014027
(2008); V. Pascalutsa, M. Vanderhaeghen, S. N. Yang, Phys. Rept. \textbf{437}, 125 (2007). 

\bibitem{Gegelia(1999):EOMS}J. Gegelia, G. Japaridze, Phys. Rev.
D \textbf{60}, 114038 (1999); J. Gegelia, G. Japaridze, X.Q. Wang, J. Phys. G \textbf{29}, 2303
(2003); T. Fuchs, J. Gegelia, G. Japaridze, S. Scherer, Phys. Rev. D \textbf{68},
056005 (2003). 

\bibitem{Alexandrou(2009):LatticeDelta} C. Alexandrou \textit{et
al.}, Nucl. Phys. A \textbf{825}, 115 (2009); Phys. Rev. D \textbf{79}, 014507
(2009).

\bibitem{Yamazaki} T. Yamazaki \textit{et al.} (RBC and UKQCD Collab.), Phys. Rev. D {\bf79}, 114505 (2009). 

\bibitem{Lin} H.-W. Lin, K. Orginos, Phys. Rev. D {\bf79}, 074507 (2009). 

\bibitem{Syritsyn}S. N. Syritsyn \textit{et al.} (LHPC Collab.), Phys. Rev. D {\bf81}, 034507 (2010).  

\bibitem{Bratt} J. D. Bratt \textit{et al.} (LHPC Collab.), Phys. Rev. D {\bf82},
  094502 (2010).

\end{thebibliography}
\end{document}